\documentclass[a4paper,12pt]{jpconf}
\usepackage[latin9]{inputenc}
\usepackage{dvipost}
\usepackage{units}
\usepackage{textcomp}
\usepackage{amsmath}
\usepackage{amssymb}
\usepackage{graphicx}

\makeatletter

\dvipostlayout
\dvipost{osstart color push Red}
\dvipost{osend color pop}
\dvipost{cbstart color push Blue}
\dvipost{cbend color pop}

\usepackage[square,sort&compress,numbers]{natbib}
\usepackage{sidecap}

\usepackage[font=small,format=plain,labelfont=bf,up]{caption}

\makeatother

\begin{document}
\title{In search of superluminal quantum communications: recent experiments and possible improvements.}

\author{B Cocciaro$^1$, S Faetti$^2$ and L Fronzoni$^2$}

\address{$^1$ High School XXV Aprile, Via Milano 2, I-56025 Pontedera (Pisa), Italy}

\address{$^2$ Department of Physics Enrico Fermi, Largo Pontecorvo 3, I-56127 Pisa, Italy}

\ead{ b.cocciaro@comeg.it, faetti@df.unipi.it, fronzoni@df.unipi.it}
\begin{abstract}
As shown in the famous \emph{EPR} paper (Einstein, Podolsky e Rosen,
1935), Quantum Mechanics is non-local. The Bell theorem and the experiments
by Aspect and many others, ruled out the possibility of explaining
quantum correlations between entangled particles using local hidden
variables models (except for implausible combinations of loopholes).
Some authors (Bell, Eberhard, Bohm and Hiley) suggested that quantum
correlations could be due to superluminal communications (tachyons)
that propagate isotropically with velocity \emph{$v_{t}>c$} in a
preferred reference frame. For finite values of \emph{$v_{t}$}, Quantum
Mechanics and superluminal models lead to different predictions. Some
years ago a Geneva group and our group did experiments on entangled
photons to evidence possible discrepancies between experimental results
and quantum predictions. Since no discrepancy was found, these experiments
established only lower bounds for the possible tachyon velocities
\emph{$v_{t}$}. Here we propose an improved experiment that should
lead us to explore a much larger range of possible tachyon velocities
\emph{$v_{t}$} for any possible direction of velocity $\vec{V}$
of the tachyons preferred frame. 
\end{abstract}

\section{Introduction}

\begin{SCfigure}[50]
 \centering
 \includegraphics[width=0.5\textwidth]{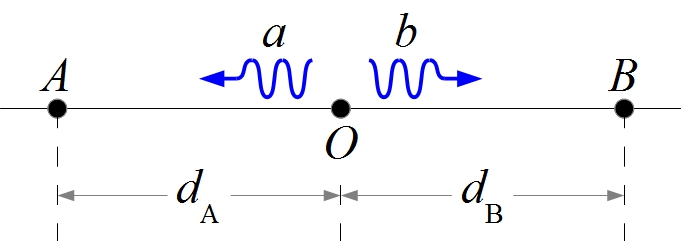}
 \hspace{0.05in}
 \caption{\textit{O}: point where a couple of entangled photons (\emph{a} and \emph{b}) are created; \textit{A} and \textit{B}: points equidistant from \textit{O} ($d_{A}=d_{B}$) where the polarization of the entangled photons is measured.}
 \label{fig:fotoni entangled}
\end{SCfigure}

The non local character of Quantum Mechanics (\textit{QM}) has been
object of a great debate starting from the famous Einstein-Podolsky-Rosen
(\textit{EPR}) paper \cite{EPR}. Consider, for instance, a quantum
system made by two photons \emph{a} and \emph{b} that are in the polarization
entangled state 

\begin{equation}
|\psi>=\frac{1}{\sqrt{2}}\left(|H,H>+e^{i\phi}|V,V>\right)\label{eq:1}
\end{equation}
where \textit{H} and \textit{V} stand for horizontal and vertical
polarization, respectively, and $\phi$ is a constant phase coefficient.
The two entangled photons are created at point \textit{O}, propagate
in space far away one from the other (see Fig.\ref{fig:fotoni entangled})
and reach at the same time points \textit{A} and \textit{B} that are
equidistant from \textit{O} as schematically drawn in Fig.\ref{fig:fotoni entangled}. 

Suppose the polarization of the two photons is measured at the same
time at points \textit{A} and \textit{B}. According to \textit{QM},
a measurement of horizontal polarization of photon \emph{a} (or \emph{b})
leads to the collapse of the entangled state to $|H,H>$, then, also
photon \emph{b} (or \emph{a}) must collapse to the horizontal polarization.
This behaviour\textit{ }suggests the existence of a sort of ``action
at a distance'' qualitatively similar to that introduced in the past
to describe interactions between either electric charges or masses.
However, according to the Maxwell electromagnetic theory and to the
Einstein General Relativity theory, it is now commonly accepted that
interactions between electric charges or masses are not instantaneous
but occur through exchange of signals (photons or gravitons). This
means that classical physical phenomena are described by local models.
Many physicists are unsatisfied of the non local character of \textit{QM}
and alternative local models have been proposed to explain quantum
correlations. The simplest way to explain the non local aspects of
\emph{QM} as well as its probabilistic behaviour is to assume a lack
of complete information about the actual state of the system in analogy
with what happens for thermodynamic systems. Then, one could explain
the probabilistic behaviour predicted by \textit{QM} as due to a not
complete knowledge of all influences affecting the system (\textit{hidden
variables}). In particular, the polarizations of the entangled photons
\emph{a} and \emph{b} would be already well defined (by their hidden
variables) when they are created at point \emph{O}. However, as shown
by Bell \cite{Bell} and other authors \cite{CHSH,Clauser_PhysRevD_1974},
the correlations between entangled particles predicted by any theory
based on local hidden variables must satisfy some inequalities that
are not satisfied by \textit{QM.} The existence of these inequalities
permits to perform experiments to decide unambiguously between hidden
variables theories and \textit{QM.} Many experiments of this kind
have been performed before and after the famous Aspect experiments
\cite{Feedman_PhysRevLett_1972,Aspect,Zeilinger_PLA_1986,Tittel_PhysRevLett_1998,Weihs_PhysRevLett_1998,Aspect_Nature_1999,Pan_Nature_2000,Grangier_Nature_2001,Rowe_Nature_2001,Matsukevich_PRL_2008}.
All the experiments (except a few old ones \cite{Faraci_LettNuovoCim_1974,Clauser_NuovoCim_1976},
see cap. 11 of \cite{Selleri_FisicaNovecento_2003} for a detailed
bibliography) demonstrated that the Bell inequalities are violated.
Although the locality loophole and the detection loophole have not
yet completely closed using a single experimental apparatus \cite{Genovese_PhysRep_2005},
experiments have separately closed both the locality loophole \cite{Aspect,Weihs_PhysRevLett_1998,Zeilinger_PLA_1986,Aspect_Nature_1999,Tittel_PhysRevLett_1998}
and the detection loophole \cite{Rowe_Nature_2001,Grangier_Nature_2001,Matsukevich_PRL_2008}.
Then, it is reasonable to think that hidden variables alone cannot
justify the experimentally observed correlations (except if implausible
combinations of loopholes are supposed to exist). For many classical
systems, correlations between two events are often explained as the
consequence of communications and, thus, also quantum correlations
between entangled particles could be due to some communication. However,
the Aspect experiments and many other \emph{EPR} experiments were
performed in space-like conditions and, thus, if Quantum correlations
would be due to communications, the communications velocity should
exceed the light velocity \emph{c}. For this reason, after the Aspect
experiment, Bell said ``\emph{in these EPR experiments there is the
suggestion that behind the scenes something is going faster than light}''\cite{Davies_ghost_1993}.
Successively, well defined models for \textit{QM} based on the presence
of superluminal communications have been proposed \cite{Eberhard_1989,Bohm_undivided_1991}.
The possibility of the existence of particles going faster than light
(\emph{tachyons}) has been proposed some years ago by some authors
\cite{Bilaniuk_AmJPh_1962,Feinberg_PhysRev_1967,Bilaniuk_PhysToday_1969}.
Tachyons are known to lead to causal paradoxes \cite{moller_theory_1955}
(the present occurring in a given point can be affected by the future
occurring in the same point). Consider, for instance, a first person
that exits from his home and is wetted by rain. He could send a tachyon
to inform a second person that send a replay tachyon that is received
by the first person before he exits from the home. Then, he could
decide to get an umbrella to be not wetted by the rain. Obviously
such a behaviour is unrealistic. However, no causal paradox arises
if tachyons are supposed to propagate in a preferred frame where the
tachyon velocity $v_{t}=\beta_{t}c$ ($\beta_{t}>1$) is the same
in all directions (see, for instance, \cite{Kowalczynski_IntJThPhys_1984,Reuse_AnPhys_1984,Caban_PhysRevA_1999,maudlin_quantum_2001,Cocciaro_riserratevi_2012}).
Note that also in this case the Special Relativity predicts that a
tachyon that is emitted at point \emph{A} at the local time $t=0$
can get point \emph{B} at time $t<0$. However, this does not represent
a paradoxical result since the time ordering between separated points
of the space has no direct physical meaning. Indeed, the time ordering
between events occurring in different space points is conventional
and depends on the conventional procedure used to synchronize far
clocks. We assume here the conventionalist thesis on the synchronization
of distant clocks \cite{Anderson_PhysRep_1998}. It is true that we
cannot yet consider closed the debate on this topic \cite{jammer_concepts_2006},
but we believe that the conventionalist thesis is the correct one.

Consider, now, two photons that are in the entangled state of Eq.\eqref{eq:1}
and suppose that photon \emph{a} passes through a polarizing filter
with horizontal polarization axis. According to the superluminal models
of \textit{QM}, when photon \emph{a} passes through the polarizing
filter, it collapses to the horizontally polarized state, then a tachyon
is sent to the entangled photon \emph{b} that collapses to the horizontally
oriented state only after this communication has been received. Therefore,
the \textit{QM} correlations between entangled photons can be recovered
only if it has been sufficient time to communicate between the two
entangled photons. Consider, for instance, an ideal experiment performed
in the tachyon preferred frame $S'$ where two polarizing filters
lie at points \emph{A} and \emph{B} at the same optical distances
$d'_{A}=d'_{B}$ from source \emph{O} of the entangled photons as
shown in Fig.\ref{fig:fotoni entangled}. In these conditions photons
\emph{a} and \emph{b} get both polarizers at the same time (in the
\emph{PF}) and, thus, if the tachyon velocity in the \emph{PF} has
a finite value, no communication is possible and correlations between
entangled particles should differ appreciably from the predictions
of \emph{QM}. Of course, there is always communication if the tachyon
velocity is $v_{t}\rightarrow\infty$, then no experiment satisfying
\textit{QM} can invalidate the superluminal models provided $v_{t}\rightarrow\infty$.
In such a case superluminal models are completely equivalent to \emph{QM}
as occurs for the Bohm model \cite{Bohm_1,Bohm_2}. It has been recently
shown \cite{Bancal_NatPhys_2012,gisin_quantum_2012} that, if \emph{QM}
correlations are due to superluminal signals with finite velocity
$v_{t}$, then superluminal signalling becomes possible, that is communications
at faster than light velocity must be possible at a macroscopic level
and they cannot be confined to ``hidden'' variables.

From the experimental point of view, equality $d'_{A}=d'_{B}$ can
be only approximatively verified within a given uncertainty $\Delta d'$.
Consequently, photons \emph{a} and \emph{b} could get the polarisers
at two different times ($\Delta t'=\nicefrac{\Delta d'}{c}$) and,
thus, they could communicate only if the tachyon velocity exceeds
a lower bound $v_{t,min}=\nicefrac{d'_{AB}}{\Delta t'}$ where $d'_{AB}$
is the distance between points \emph{A} an \emph{B} in the \emph{PF}.
The possible results of such an experiment are:\emph{ i)} a lack of
quantum correlations is observed; \emph{ii)} quantum correlations
are always satisfied. In the first case (\emph{i)}) one can conclude
that orthodox \emph{QM} is not correct and that quantum correlations
are due to exchange of superluminal messages. By suitably changing
distances $d'_{A}$ and $d'_{B}$ one could obtain a measure of the
tachyon velocity $v_{t}$. In the second case (\emph{ii)}), due to
the experimental uncertainty, one cannot invalidate the tachyon model
of \emph{QM} but can only establish a lower bound $v_{t,min}$ for
the tachyon velocity. So far we assumed that the experiment is carried
out in the preferred frame, but velocity vector \textbf{\emph{$\vec{V}$}}
of \emph{PF} is unknown and, thus, the experiment cannot be performed
in this frame. It will be shown in Section \ref{sec:The-main-features}
that this drawback can be bypassed performing the experiment on the
Earth with the \emph{A}-\emph{B} axis aligned along the West-East
direction. Such an experiment could provide both the tachyon velocity
$v_{t}$ and the the velocity vector \textbf{\emph{$\vec{V}$}} of
the \emph{PF}.

A long-distance (10.6 km) \textit{EPR} experiment to detect possible
effects of superluminal quantum communications has been performed
by Scarani et al. \cite{scarani_PLA_2000} using energy-time entangled
photons. No deviation from the predictions of \emph{QM} was observed
and, thus, the authors obtained only a lower bound for the tachyon
velocities in the \emph{PF}. The experimental results were analyzed
under the assumption that the preferred frame is the frame of cosmic
microwave background radiation. With this assumption, the authors
obtained a lower bound $v_{t,min}=1.5\times10^{4}\, c$. Successively,
similar long-distance measurements have been performed by Salart et
al. \cite{Salart_nature_2008} improving some features of the previous
experiment and using detectors aligned close to West-East direction
(at angle $\alpha=5.8\text{\textdegree}$). The authors found a lower
bound for the tachyon velocity for many different possible directions
of velocity \textbf{\emph{$\vec{V}$}} of the \emph{PF}. More recently
\cite{Cocciaro_PLA_2011} we performed \emph{EPR} measurements on
polarization entangled photons in a laboratory experiment with small
distances $d_{A}\thickapprox d_{B}\thickapprox1\,\mathrm{m}$ and
with the \emph{AB} axis in Fig.\ref{fig:fotoni entangled} precisely
aligned along the West-East direction ($\left|\alpha\right|<0.2\text{\textdegree}$).
In our experiment, too, no deviation from the predictions of \emph{QM}
was found and, thus, we obtained only a lower bound for the tachyon
velocity. The choice of aligning the measurement points \emph{A} and
\emph{B} just along the West-East direction allowed us to obtain a
lower bound for the tachyon velocity for any possible orientation
of the velocity of the preferred frame. The experiment in \cite{Salart_nature_2008}
and in \cite{Cocciaro_PLA_2011} are somewhat complementary. Indeed,
the Salart et al. experiment used very large distances ($d_{A}\thickapprox d_{B}\thickapprox10\,\mathrm{km}$)
but somewhat large acquisition times ($\delta t\thickapprox360\,\mathrm{s}$),
while our experiment used shorter distances ($d_{A}\thickapprox d_{B}\thickapprox1\,\mathrm{m}$)
but much shorter acquisition times ($\delta t\thickapprox4\,\mathrm{s}$).
With these features, the Salart et al. experiment was much more sensitive
to tachyons propagating in \emph{PF} that have velocities much smaller
than the light velocity whilst our experiment was more sensitive to
\emph{PF} travelling at higher velocities (see Fig.\ref{fig:4}).
As will be shown in Section \ref{sec:The-main-features}, an \emph{EPR}
experiment similar to our previous experiment but with entangled photons
that propagate in air at much larger distances (of order of 1 km)
and with much smaller acquisition times (of order of 0.1 s) could
increase greatly the range of detectable tachyon velocities. To get
this goal, we plan to perform such an experiment inside the long galleries
of the \emph{EGO} structure (European Gravitational Observatory) \cite{EGO}.
In this paper, we will analyse the main features of the proposed experiment.
The possible results of this experiment are either the detection of
possible discrepancies between experiment and \emph{QM} due to a finite
velocity superluminal communication or the increase by about two orders
of magnitude of the actual lower bounds for the tachyon velocities.

The main features of the experiment are discussed in Section \ref{sec:The-main-features}.
The critical points and the main sources of experimental uncertainty
are discussed in Section \ref{sec:Critical-points-and}. Finally,
the conclusions are given in Section \ref{sec:Conclusions}.

\section{\label{sec:The-main-features}The main features of the experimental
method.}

Consider the geometry of Fig.\ref{fig:fotoni entangled}. We start
considering the ideal case where the experiment is performed in the
preferred frame \emph{$S'$ }with tachyons that propagate with the
same velocity $v_{t}=\beta_{t}c$ ($\beta_{t}>1$) along any direction
\cite{Salart_nature_2008}. For simplicity, in the following, we will
call ``tachyon velocity'' the reduced velocity $\beta_{t}=\frac{v_{t}}{c}$.
Two polarizing filters lie at points \emph{A} and \emph{B} aligned
along a \emph{x}'-axis and at optical distances $d'_{A}$ and $d'_{B}$
from the source \emph{O} of the entangled photons (the apostrophe
denotes the parameters measured in the \emph{PF}). The two entangled
photons will get both the polarisers at the same times $t'_{A}=t'_{B}$
if $d'_{A}=d'_{B}$ and, thus, no superluminal communication can be
possible if the tachyon velocity has a finite value. Due to the experimental
uncertainty $\Delta d'$, distances $d'_{A}$ and $d'_{B}$ can never
be exactly equalized and $\left|\Delta t'\right|=\left|t'_{A}-t'_{B}\right|=\nicefrac{\Delta d'}{c}\neq0$.
In these conditions, a superluminal communication between points \emph{A}
and \emph{B} will be not possible only if the time that a tachyon
spend to go from \emph{A} to \emph{B} (or from \emph{B} to \emph{A})
is greater than $\left|\Delta t'\right|$, that is if the tachyon
velocity $\beta_{t}$ is smaller than 

\begin{equation}
\beta_{t,min}=\frac{d'_{AB}}{\left|\Delta ct'\right|},\label{eq:2}
\end{equation}
where\emph{ }$d'_{AB}$ is the distance between points \emph{A} and
\emph{B. QM }correlations will be always established if the tachyon
velocity $\beta_{t}$ is higher than $\beta_{t,min}$ of Eq.\eqref{eq:2}
but appreciable differences between \emph{QM} correlations and experimentally
observed correlations is expected if $\beta_{t}<\beta_{t,min}$. In
this latter case, the experimental correlations should satisfy the
Bell inequality. Therefore, to detect possible discrepancies between
experiment and \emph{QM} due to the finite velocity of tachyons, one
has to increase the lower limit in Eq.\eqref{eq:2} as much as possible
to satisfy condition $\beta_{t}<\beta_{t,min}$. This goal could be
achieved either reducing $\Delta d'=\left|\Delta ct'\right|$ or increasing
distance $d'_{AB}$ between the polarisers. If discrepancies between
\emph{QM} and experiment would be found for given values of $\Delta d'$
and $d'_{AB}$, one could conclude that the tachyon velocity is lower
than the lower bound in Eq.\eqref{eq:2}. In such a case, one could
measure the tachyon velocity $\beta_{t}$ changing distances $d'_{A}$,
$d'_{B}$ and $d'_{AB}$ to reduce $\beta_{t,min}$ of Eq.\eqref{eq:2}
until condition $\beta_{t,min}<\beta_{t}$ is verified and\emph{ QM}
correlations are re-established. The tachyon velocity $\beta_{t}$
would correspond to this critical value of $\beta_{t,min}$. 

\begin{figure}
\begin{centering}
\includegraphics[scale=0.3]{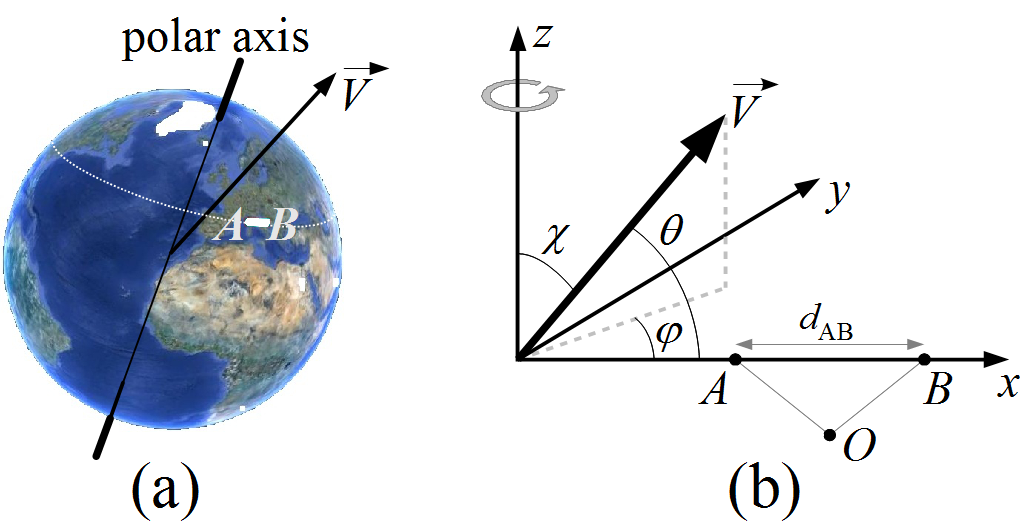}
\par\end{centering}

\caption{\label{fig:2}(a) geometry of the experiment with segment \emph{AB}
oriented along the West-East direction on the Earth; (b) detail of
the geometric parameters that characterize the experiment.\textbf{\emph{
}}\emph{z} is the North-South axis and \emph{x} is the West-East axis.\textbf{\emph{
$\vec{V}$}} is the velocity vector of the \emph{PF} with respect
to the laboratory,\emph{ }\textbf{\emph{$\theta$}} denotes the angle
of \textbf{\emph{$\vec{V}$}} with the \emph{x}-axis, \textbf{\emph{$\varphi$}}
is the azimuthal angle and \textbf{\emph{$\chi$}} is the polar angle.
The polarizing filters that collect the entangled photons lie at points
\emph{A} and \emph{B} aligned along the \emph{x}-axis at the same
optical distances from the source \emph{O} of the entangled photons.}
\end{figure}

\begin{SCfigure}[50]
 \centering
 \includegraphics[width=0.4\textwidth]{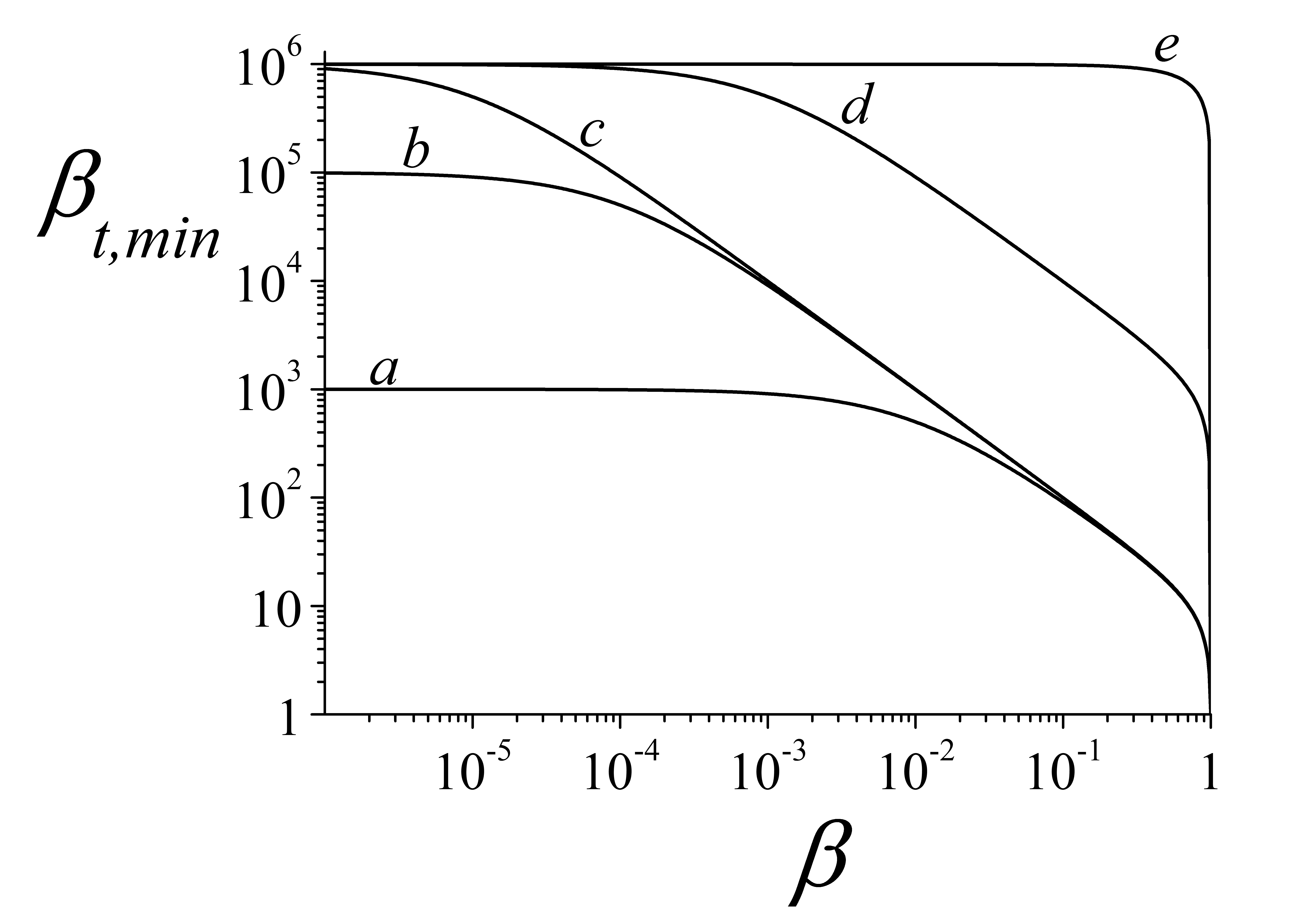}
 \hspace{0.05in}
 \caption{The typical plot of function $\beta_{t,min}$ versus $\beta$ is shown for the unfavourable case $\chi=\nicefrac{\pi}{2}$ and for some values of the experimental parameters $\bar{\rho}$ and $\delta t$. Curves \emph{a}, \emph{b} and \emph{c} correspond to the fixed acquisition time $\delta t=10^{-1}\times\frac{T}{\pi}$ and to decreasing values of $\bar{\rho}$ ($a:\bar{\rho}=10^{-3},\, b:\bar{\rho}=10^{-5},\, c:\bar{\rho}=10^{-6}$). Curves \emph{c}, \emph{d} and \emph{e} correspond to a fixed value $\bar{\rho}=10^{-6}$ and to decreasing values of $\delta t$ ($c:\delta t=10^{-1}\times\frac{T}{\pi},\, d:\delta t=10^{-3}\times\frac{T}{\pi},\, e:\delta t=10^{-7}\times\frac{T}{\pi}$). Note that curve \emph{e} satisfies condition \eqref{eq:7} and, thus, it depends on $\beta$ only for \emph{PF} moving at improbable relativistic velocities.}
 \label{fig:3}
\end{SCfigure}

The results above were obtained assuming that the experiment is performed
in the tachyons \emph{PF}. Of course, this is not possible because
we do not know what is the direction and the magnitude of the \emph{PF}
velocity $\vec{V}=\vec{\beta}c$. However, this apparent difficulty
can be overcome if one takes advantage of the rotation motion of the
Earth around its axis and aligns points \emph{A} and \emph{B} along
the West-East \emph{x}-axis on the Earth as shown in Fig.\ref{fig:2}.
Suppose that points \emph{A} and \emph{B} are precisely aligned along
the \emph{x}-axis and that optical distances $d{}_{A}$ and $d{}_{B}$
of \emph{A} and \emph{B} from the entangled photon source \emph{O}
have the same values on the Earth frame. In this condition, the two
entangled photons get the two polarizing filters at the same times
\textbf{\emph{$t_{A}$}} and\textbf{\emph{ $t_{B}$}} in the Earth
frame but the arrival times \textbf{\emph{$t'_{A}$}} and\textbf{\emph{
$t'_{B}$}} can be different in the tachyon \emph{PF}. Due to the
rotation of the Earth with the angular velocity \textbf{\emph{$\omega$}},
angle \textbf{\emph{$\theta$}} between the the velocity \textbf{\emph{$\vec{V}$}}
of the \emph{PF} and the\emph{ x}-axis (West-East direction) changes
periodically with time \emph{t} according to the simple law:

\begin{equation}
\theta(t)=\arccos\left[\sin\chi\cos\omega\left(t-t_{0}\right)\right],\label{eq:3}
\end{equation}
where $t_{0}$ is the unknown time which gives $\varphi\left(t\right)=0$
in Fig.\ref{fig:2}. Angle \textbf{\emph{$\theta$}} oscillates periodically
between a minimum value \textbf{\emph{$\nicefrac{\pi}{2}-\chi$}}
(\textbf{\emph{$\varphi(t)$}}=0 in Fig.\ref{fig:2}) and a maximum
value \textbf{\emph{$\nicefrac{\pi}{2}+\chi$}} (\textbf{\emph{$\varphi(t)$}}=\textbf{\emph{$\pi$}}
in Fig.\ref{fig:2}). Then, whatever is the orientation of the velocity
vector \textbf{\emph{$\vec{V}$}} of the \emph{PF}, there are two
times \textbf{\emph{$t_{1}$}} and\textbf{\emph{ $t_{2}$}} during
each sidereal day where \textbf{\emph{$\vec{V}$}} becomes perpendicular
to the West-East \emph{x}-axis. At these two times, according to the
Special Relativity, the distances of \emph{A} and \emph{B} from \emph{O}
are equal also in the \emph{PF} frame ($d'_{A}=d'_{B}$) and, thus,
the arrival of the entangled photons at points \emph{A} and \emph{B}
is simultaneous also in the \emph{PF}. Then, deviations of correlations
from the predictions of the\emph{ QM} should be observed at the special
times \textbf{\emph{$t_{1}$}} and\textbf{\emph{ $t_{2}$}} if the
optical paths of the entangled photons would be equal. Of course,
due to the experimental uncertainty $\Delta d$ on the equalization
of the optical paths, deviations from the predictions of \emph{QM
}could be only observed if the tachyon velocity $\beta_{t}$ is lower
than a lower bound $\beta_{t,min}$. 
Furthermore, the need of performing the experiment in the Earth frame
introduces also another source of uncertainty because vector \textbf{\emph{$\vec{V}$}}
becomes orthogonal to the West-East axis only at two well defined
times \textbf{\emph{$t_{1}$}} and\textbf{\emph{ $t_{2}$}}. However,
to detect a statistically significant number of coincidences of entangled
photons, a sufficiently long acquisition time $\delta t$ is needed.
Also if this acquisition time is centred around the special times
\textbf{\emph{$t_{1}$}} and\textbf{\emph{ $t_{2}$}}, the velocity
vector \textbf{\emph{$\vec{V}$}} does not remain exactly perpendicular
to the \emph{x}-axis during the whole acquisition time. Therefore,
the finite acquisition time leads to an adjunctive uncertainty on
the equality of the optical paths with a consequent decreasing of
the value of $\beta_{t,min}$. In conclusion, two main parameters
determine the lower limit of $\beta_{t,min}$ obtainable with the
measurements on the Earth frame: 1) the accuracy $\Delta d$ on the
equalization of the optical paths; 2) the finite acquisition time
$\delta t$. The lower bound $\beta_{t,min}$, defined by Eq.\eqref{eq:2},
can be rewritten in terms of physical parameters measured in the Earth
frame and becomes \cite{Salart_nature_2008,Cocciaro_PLA_2011}:

\begin{equation}
\beta_{t,min}=\sqrt{1+\frac{\left(1-\beta^{2}\right)\left[1-\bar{\rho}^{2}\right]}{\left[\bar{\rho}+\beta\sin\chi\sin\frac{\pi\delta t}{T}\right]^{2}}},\label{eq:4}
\end{equation}
where $\bar{\rho}=\frac{\Delta d}{d_{AB}}$, \emph{T} is the sidereal
day, $\delta t$ is the acquisition time, $\chi$ is the polar angle
between the North-South axis of the Earth and velocity \textbf{\emph{$\vec{V}$}}
of the \emph{PF }(see Fig.\ref{fig:2}) and $\beta$ is the reduced
velocity ($\beta=\nicefrac{V}{c}$) of the \emph{PF}. A quick way
to obtain \eqref{eq:4} is shown in the Appendix. In typical experimental
conditions \cite{Salart_nature_2008,Cocciaro_PLA_2011}, the acquisition
time $\delta t$ is much smaller than the sidereal day \emph{T }and
$\beta_{t,min}$ is a decreasing function of both $\bar{\rho}$ and
$\delta t$ and reaches a minimum value if $\chi=\nicefrac{\pi}{2}$.
Our following considerations and figures will be restricted to $\delta t\ll T$
and to the unfavourable condition $\chi=\nicefrac{\pi}{2}$. $\beta_{t,min}$
is also a decreasing function of $\beta$ that assumes its maximum
value $\beta_{t,min}=\frac{1}{\bar{\rho}}$ for $\beta=0$ and approaches
the minimum value $\beta_{t,min}=1$ for $\beta\rightarrow1$. The
typical plot of function $\beta_{t,min}$ versus the reduced velocity
$\beta$ of the \emph{PF} for $\chi=\nicefrac{\pi}{2}$ is drawn in
Fig.\ref{fig:3} for some values of the experimental parameters $\bar{\rho}$ and
$\delta t$. $\beta_{t,min}$ keeps an almost constant value $\beta_{t,min}\thickapprox\frac{1}{\bar{\rho}}$
for $\beta\ll\beta_{0}=\nicefrac{\bar{\rho}T}{(\pi\delta t)}$ , then
it decreases rapidly as $\beta$ increases above $\beta_{0}$. Eq.\eqref{eq:4}
and Fig.\ref{fig:3} evidence what is the optimal strategy to increase
$\beta_{t,min}$ as much as possible. First of all we must reduce
$\bar{\rho}$ minimizing the uncertainty $\Delta d$ and increasing
distance $d_{AB}$ as far as possible. Then, we must use an acquisition
time $\delta t$ sufficiently small to render negligible the $\delta t$-contribution
in Eq.\eqref{eq:4}. This contribution is always negligible if:

\begin{equation}
\delta t\ll\bar{\rho}\frac{T}{\pi}.\label{eq:5}
\end{equation}
If condition \eqref{eq:5} is satisfied, the lower limit $\beta_{t,min}$
becomes insensitive to the acquisition time $\delta t$.

\begin{SCfigure}[50]
 \centering
 \includegraphics[width=0.4\textwidth]{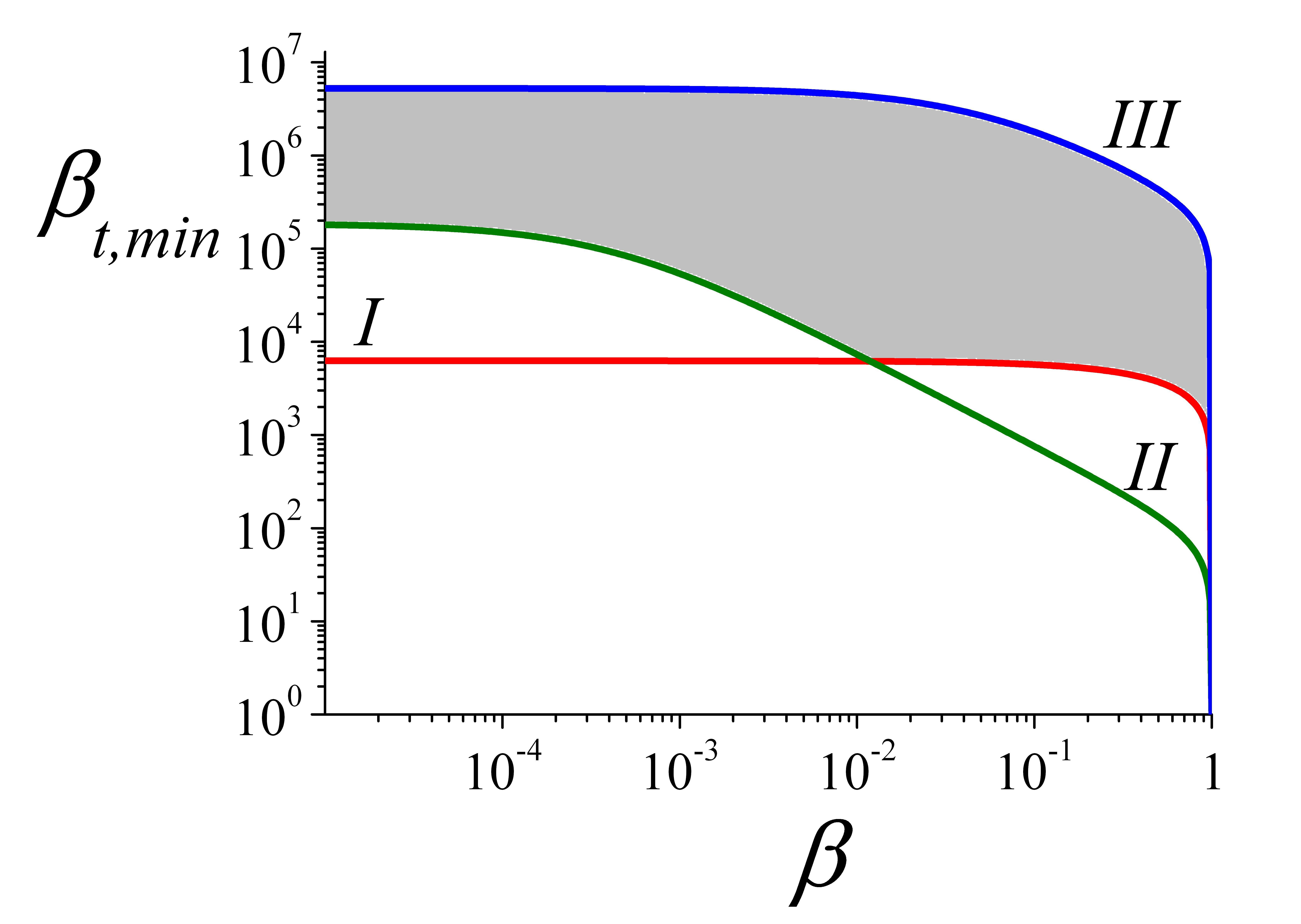}
 \hspace{0.05in}
 \caption{The three curves \emph{I} (red), \emph{II} (green) and \emph{III} (blue) correspond to the experimental results for the lower bound $\beta_{t,min}$ versus $\beta$ in different experiments. \emph{I}: our previous results ($\bar{\rho}=1.6\times10^{-4}$, $\delta t=4 $s) ; \emph{II}: the Geneva group results ($\bar{\rho}=5.4\times10^{-6}$, $\delta t=360 $s) ; \emph{III}: the predicted results for the experiment proposed here ($\bar{\rho}=1.9\times10^{-7}$, $\delta t=0.1 $s). The gray region represents the new region of tachyons velocities $\beta_{t}$ that would become accessible with the new experiment.}
 \label{fig:4}
\end{SCfigure}

In the Geneva experiment \cite{Salart_nature_2008} the very small
value $\bar{\rho}=5.4\times10^{-6}$ was obtained using Telecom optical
fibres connecting two Telecom stations at a distance of about $20\,\mathrm{km}$
and aligned approximatively along the West-East direction. The experiment
was performed using energy-time entangled photons. The two Telecom
stations (\emph{A} and \emph{B}) were not exactly aligned along the
West-East axis but did the angle $\gamma=5.8\text{\textdegree}$ with
this axis and, thus, the experiment was poorly sensitive to tachyons
\emph{PF} travelling with velocity \textbf{\emph{$\vec{V}$}} lying
in a cone of aperture $\approx6\text{\textdegree}$ around the North
south axis. Indeed, due to the Earth rotation, the angle between the
\emph{AB} axis and the velocity vector \textbf{\emph{$\vec{V}$}}
oscillates with time between the minimum value $\nicefrac{\pi}{2}-\chi-\gamma$
and the maximum value $\nicefrac{\pi}{2}+\chi-\gamma$. Then, the
\emph{AB} axis can become orthogonal to \textbf{\emph{$\vec{V}$}}
only if $\chi\geq\gamma$. In this long path experiment the main effect
that limited a further reduction of $\bar{\rho}$ was due to the appreciable
optical dispersion of the photon wave-packet in the long Telecom fibres.
On the other hand, due to losses in the optical fibres, the photons
count rate was relatively small and a somewhat long measurement time
($\delta t\approx360\,\mathrm{s}$) was needed to obtain a statistically
significant number of coincidences. In these conditions the experiment
was greatly sensitive to tachyons \emph{PF} travelling at speeds much
smaller than that the light velocity and much less sensitive to relativistic
\emph{PF} (see curve\emph{ II} in Fig.\ref{fig:4}). The Pisa experiment
\cite{Cocciaro_PLA_2011} was performed with entangled photons that
propagate in air over much shorter distances (about $2\,\mathrm{m}$)
and the polarization correlations were measured instead of energy-time
correlations. In these conditions photon losses were minimized and
the typical acquisition time was $\delta t\approx4\,\mathrm{s}$.
An interferometric method was used to minimize the uncertainty on
the equalization of the photons optical paths and the main source
of uncertainty $\Delta d$ was due to the $220\,\mu\mathrm{m}$ thickness
of the absorbing layer of the polarizing filters. The obtained value
of parameter $\bar{\rho}$ was $\bar{\rho}=1.6\times10^{-4}$ that
is about 30 times higher than the value characterizing the experiment
of \cite{Salart_nature_2008}. Then, our experiment is much less sensitive
than that in \cite{Salart_nature_2008} to tachyons propagating in
a low speed \emph{PF} but is more sensitive to tachyons propagating
in high speed \emph{PF} (see curve\emph{ I} in Fig.\ref{fig:4}).
In this sense, our experiment was somewhat complementary to that of
the Geneva Group. In the present paper we propose a new experiment
exploiting the propagation of polarized entangled photons in air at
distances about 850 times longer than in our previous experiment.
The experiment should be performed inside the long tunnels of the
European Gravitational Observatory (\emph{EGO}) that host the \emph{VIRGO}
experiment on the detection of gravitational waves \cite{VIRGO}.
Using an interferometric method and feedback procedures we will hold
the uncertainty $\Delta d$ on the equality of the optical paths of
the entangled photons well below the basic uncertainty due to the
thickness of the absorbing layer of the polarizing filters (220 $\mu$m).
The small uncertainty and the long optical paths will allow us to
decrease our previous value $\bar{\rho}=1.6\times10^{-4}$ up to the
much smaller value $\bar{\rho}=1.9\times10^{-7}$. As shown in the
discussion above, the lower bound for the tachyon velocities is also
determined by the acquisition time $\delta t$ that must satisfy the
condition that the number of detected coincidences must be statistically
relevant. $\delta t$ can be reduced increasing the brightness of
the source of the entangled photons and minimizing the losses. In
some papers, the Kwiat group \cite{Kwiat_OptExpr_2005,Kwiat_OptExpr_2007,Kwiat_OptExpr_2009}
developed a very efficient and simple method to obtain highly bright
sources of entangled photons with a high degree of entanglement. Using
this technique and a suitable optical configuration we plan to increase
by more than 1000 times the number of measured coincidences and to
reduce the acquisition time to less than $\delta t=0.1\,\mathrm{s}$.
The lower bounds obtained in the previous experiments together with
the new lower bound that should be reached with the experiment proposed
here are shown in Fig.\ref{fig:4}. Curve \emph{I} (red) corresponds to our previous experimental results,
curve \emph{II} (green) to the results of the Geneva group and curve
\emph{III} (blue) to the results expected with the proposed experiment.
The grey region corresponds to the new region of tachyon velocities
that could become accessible.

In the next section, the critical points of the proposed experiment
and the main sources of experimental uncertainty will be discussed.

\section{\label{sec:Critical-points-and}The experiment: critical points and
main sources of experimental uncertainty}

\begin{SCfigure}[50]
 \centering
 \includegraphics[width=0.4\textwidth]{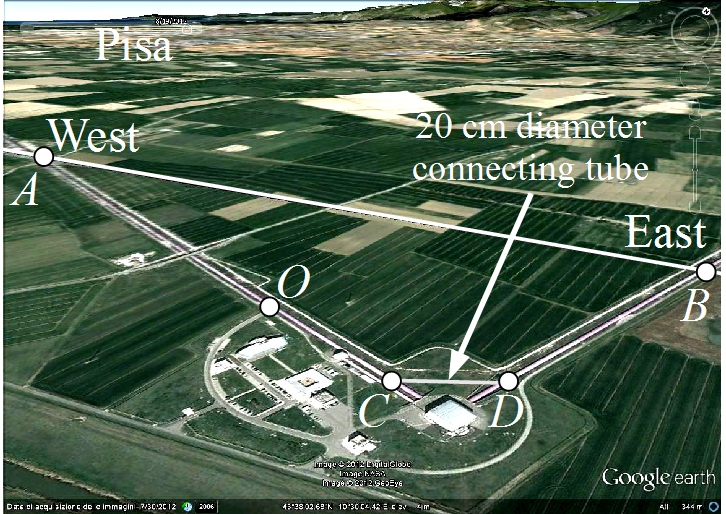}
 \hspace{0.05in}
 \caption{ A Google earth view of the \emph{EGO} structure with the two orthogonal tunnels is shown together with the West-East line. Point \emph{O} represents the position of the source of entangled photons, while \emph{A} and \emph{B} denote the positions of the two polarizing filters in the two \emph{EGO} arms. The full line represents the West-East direction. The entangled photon that travel at the right of point \emph{O} will be deviated from the tunnel containing source \emph{O} to the other\emph{ }tunnel passing through a suitable \emph{CD} tube (20 cm diameter) that will be built to connect the two arms of the \emph{EGO} structure.}
 \label{fig:5}
\end{SCfigure}

As stated above, the polarizing filters should be aligned along the
West-East direction, but the \emph{EGO} tunnels are oriented at 19\textdegree{}
and 109\textdegree{} with respect to this direction, respectively.
Therefore, the polarizing filters must be placed in two different
tunnels and one of the entangled photons must be deviated from a \emph{EGO}
tunnel to the other through the small \emph{CD }tube ($20\,\mathrm{cm}$
diameter, $100\,\mathrm{m}$ length) built to connect them. The source
of the entangled photons (point \emph{O}) will be placed in the 19\textdegree{}
tunnel together with polariser $P_{A}$ at distance $\approx800\,\mathrm{m}$
while polariser $P_{B}$ will be placed at point \emph{B} in the other
\emph{EGO} tunnel at the same distance from \emph{O} as shown schematically
in Fig.\ref{fig:5} ($\overline{OA}=\overline{OC}+\overline{CD}+\overline{DB}$).

\begin{figure}
\centering{}\includegraphics[scale=0.3]{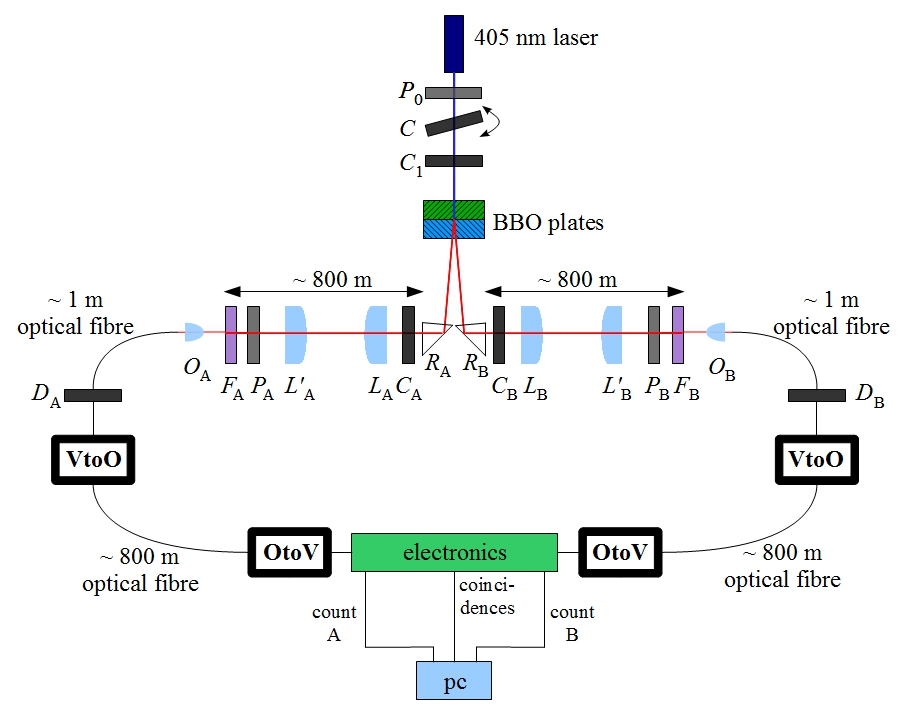}\caption{\label{fig:6} Schematic view of the experimental apparatus. To simplify
the drawing, the optical path followed by the left entangled photon
has been represented by a straight line and the optical prisms that
deviate the beam from a tunnel to the other (see Fig.\ref{fig:5})
are not shown. A $405\,\mathrm{nm}$ laser beam impinges on two adjacent
\emph{BBO} plates with orthogonal optical axes and produces two 810
nm entangled beams preferentially emitted at the symmetric angles
$\alpha_{A}=-\alpha_{B}\approx2\text{\textdegree}$. $C,\, C{}_{1},\, C_{A}$
and $C_{B}$ are optical compensators, $R_{A}$ and $R_{B}$ are two
right angle prisms; $L_{A},\, L'_{A},\, L_{B},\, L'_{B}$ are large
diameter ($18\,\mathrm{cm}$) planoconvex lenses, $P_{A}$ and $P_{B}$
are polarizing filters that are at the same distance from source \emph{O},
$F_{A}$ and $F_{B}$ are optical interferometric filters with 10
nm line-width, $O_{A}$ and $O_{B}$ are aspheric lenses and $D_{A}$
and $D_{B}$ are single photons counters. \emph{V}to\emph{O} and \emph{O}to\emph{V}
are voltage pulses to optical pulses converters and vice-versa.}
\end{figure}

The experimental apparatus is schematically shown in Fig.\ref{fig:6}
 where, for simplicity of drawing, the path \emph{OCDB } shown in
Fig.\ref{fig:5} has been represented by a straight line and the prisms
needed to deviate the photons from one \emph{EGO} arm to the other
are not represented. A blue diode laser beam ($\lambda=405\,\mathrm{nm}$)
is polarized at 45\textdegree{} with respect to the vertical axis
by polariser $P_{0}$, passes through a tilting plate optical compensator
(\textit{C}) with vertical extraordinary axis and the quartz compensator
($C_{1}$) with horizontal extraordinary axis and, finally, impinges
at normal incidence on two thin ($0.5\,\mathrm{mm}$) adjacent non-linear
optical crystals (\textit{BBO}) cut for type-I phase matching \cite{Kwiat_PhysRevA_1999}.
The optic axes of the adjacent \emph{BBO} crystals are tilted at the
angle 29.2\textdegree{} and lie in planes perpendicular to each other
with the first plane that is horizontal. The pump beam induces down
conversion at $\lambda=810\,\mathrm{nm}$ in each crystal \cite{Kwiat_PhysRevA_1999}
with a maximum of emission at the two symmetric angles $\alpha_{A}=-\alpha_{B}\approx2\text{\textdegree}$
with respect to the pump laser beam. The down converted photons are
created in the maximally entangled state $\left(|H,H>+e^{i\phi}|V,V>\right)/\sqrt{2}$,
where phase $\phi$ can be adjusted tilting the optical compensator
\textit{C}. $R_{A}$ and $R_{B}$ are two right angle prisms, $C_{A}$
and $C_{B}$ are optical \emph{BBO} plates with tilt angle 29.2\textdegree{},
$P_{A}$ and $P_{B}$ are thin near infrared polarizing films (LPNIR,
Thorlabs), $F_{A}$ and $F_{B}$ are interference filters ($\lambda=810\,\mathrm{nm}\pm5\,\mathrm{nm}$)
and $D_{A}$ and $D_{B}$ are single photons counters (Perkin Elmer
SPCM-AQ4C). $L_{A},L_{B},L'_{A}$ and $L'_{B}$ are large plano-convex
optical lenses (18 cm-diameter) that ensure that all the entangled
photons emitted within a cone of $\approx0.35\text{\textdegree}$
aperture angle (around the two maximum emission directions $\alpha_{A}=-\alpha_{B}\approx2\text{\textdegree}$)
can be collected by aspheric objectives $O_{A}$ and $O_{B}$ and
sent to detectors $D_{A}$ and $D_{B}$. The centres of polarisers
$P_{A}$ and $P_{B}$ are aligned along a \textit{x}-axis in the West-East
direction within $\approx0.1\text{\textdegree}$. The role of $P_{0}$,
\emph{C}, $P_{A}$, $P_{B}$, $F_{A}$ and $F_{B}$ is shown in \cite{Kwiat_PhysRevA_1999};
the role of $C_{1}$, $C_{A}$ and $C_{B}$ is explained in \cite{Kwiat_OptExpr_2005,Kwiat_OptExpr_2007,Kwiat_OptExpr_2009}
and is resumed in Section \ref{sub:Minimization-of-the}. The outputs
electric pulses of the detectors ($20\,\mathrm{ns}$ width) are transformed
into optical pulses, sent via optical fibres to the central region
close to the photon source, transformed again to electric pulses and
sent to electronic counters and to a coincidence circuit connected
to a \textit{PC}. A labview program controls any experimental feature. 

The results shown in curve \emph{III} of Fig.\ref{fig:4} were obtained
assuming that the uncertainty $\Delta d$ can be maintained much lower
than the thickness of the polarizing layers ($\approx220\,\mu\mathrm{m}$)
and the acquisition time can be small enough ($\delta t\approx0.1\,\mathrm{s}$).
Here below we will describe how both these conditions can be satisfied.

\subsection{\label{sub:Minimization-of-the}Minimization of the acquisition time
$\delta t$.}

The minimum acquisition time $\delta t$ is determined by the condition
that the number of detected entangled photons must be statistically
significant during time $\delta t$ ($>1000$ measured coincidences).
Then, low values of $\delta t$ can be obtained only if: \emph{a})
the number of entangled photons collected by lenses $L_{A}$ and $L_{B}$
is large enough, \emph{b}) the losses of the entangled photons in
the path from the source to the detectors are as smaller as possible.
The momentum conservation would require that the entangled photons
are emitted at two well defined angles ($\alpha_{A}=-\alpha_{B}\approx2\text{\textdegree}$),
but, due to the finite thickness of the \emph{BBO} plates ($0.5\,\mathrm{mm}$),
a $\approx\pm0.35\text{\textdegree}$ spreading around these directions
occurs. Then, a large fraction of the entangled photons can be collected
only if the aperture angle of lenses $L_{A}$ and $L_{B}$ with respect
to the photon source is $\approx0.35\text{\textdegree}$. However,
due to the great optical anisotropy of the \emph{BBO} plates, the
relative phase between entangled photons ($\phi$ in Eq.\eqref{eq:1})
depends greatly on the emission direction. This means that the entanglement
of the photons is satisfactory (close to 100\% fidelity) only if the
aperture angle is $\ll0.05\text{\textdegree}$. Furthermore, the number
of entangled photons greatly increases with the power of the pump
laser diode and, thus, a high power laser diode should be used. However,
high power laser diodes ($>50\,\mathrm{mW}$) are characterized by
a small coherence length which is typically $\approx0.2\,\mathrm{mm}$.
In these conditions, the \emph{V}-polarized down converted photons
that are produced in the first \emph{BBO} plate ($0.5\,\mathrm{mm}$
thick) are essentially uncorrelated with respect to the \emph{H}-polarized
photons produced in the second adjacent plate and, thus, the entanglement
is very poor. This latter drawback could be avoided using high coherence
length laser diodes but, in this case, the laser power would be smaller
than $50\,\mathrm{mW}$. The Kwiat group \cite{Kwiat_OptExpr_2005,Kwiat_OptExpr_2007,Kwiat_OptExpr_2009}
showed that the drawbacks described above can be bypassed using suitable
optical compensators ($C_{1}$, $C_{A}$ and $C_{B}$ in Fig.\ref{fig:6}).
$C_{1}$ is a quartz plate that introduces a retardation between the
\emph{V} and the \emph{H} components of the pump beam virtually equal
to the difference between the emission times of down converted photons
from the two \emph{BBO} plates. In this way, the losses of entanglement
due to the low coherence of the laser beam are virtually eliminated.
$C_{A}$ and $C_{B}$ are two anisotropic plates (\emph{BBO}) with
a suitable thickness to compensate the phase differences between entangled
photons propagating along different directions. Using these compensation
methods with a $280\,\mathrm{mW}$ laser beam and collecting entangled
photons with a 0.35\textdegree{} aperture angle, the Kwiat group obtained
an ultra bright and high fidelity (99 \%) source of entangled photons
with $1.02\times10^{6}$ detected coincidences/s. In our experiment
we require also that all the entangled photons that are emitted within
the 0.35\textdegree{} aperture angle get photodetectors $D_{A}$ and
$D_{B}$ at a great distance ($\approx800\,\mathrm{m}$) from the
source. Furthermore, it is convenient to maintain the diameter of
the entangled beam sufficiently smaller when passes through the 20
cm-diameter connecting tube. Both these conditions are satisfied using
large diameter ($18\,\mathrm{cm}$) plano-convex lenses with long
focal length $f=10\,\mathrm{m}$ and slightly focusing the pump beam
on the \emph{BBO} plates to have a source of entangled photons with
a small diameter \emph{D} ($D\lesssim0.5\,\mathrm{mm}$). Lenses $L_{A}$
and $L_{B}$ are put at a distance from the source slightly higher
than \emph{f }to produce\emph{ }real images of the source\emph{ }approximatively
at distance $d_{i}\approx400\,\mathrm{m}$ from $L_{A}$ and $L_{B}$.
In this condition, the image produced by lens $L_{B}$ occurs in the
centre of the connecting tube. The image diameter is $D_{i}\approx D\nicefrac{d_{i}}{f}=$2
cm that is much smaller than the $20\,\mathrm{cm}$-diameter of the
connecting tube. Furthermore, due to the large diameter of lenses,
the enlargement produced by diffraction is negligible. In these conditions
all the entangled photons emitted in the cones of aperture angle 0.35\textdegree{}
are collected by the large diameter lenses $L_{A},\, L'_{A}$ and
$L_{B},\, L'_{B}$ and real images of the source with diameter $D'_{i}\approx D=0.5\,\mathrm{mm}$
are generated on the surfaces of polarisers $P{}_{A}$ and $P{}_{B}$
at distance slightly higher than $f=10\,\mathrm{m}$ from lenses $L'_{A}$
and $L'_{B}$. The optical rays exiting from these images are collected
by the aspheric objectives $O_{A}$ and $O{}_{B}$ that focus the
entangled photons on two $60\,\mu\mathrm{m}$-diameter optical fibres
connected to detectors $D_{A}$ and $D{}_{B}$. Using the Zemax optical
design program we have verified that this geometric arrangement ensures
that all the entangled photons emitted by the source within the 0.35\textdegree{}
aperture angle are collected by detectors.

The previous analysis was performed disregarding any effect due to
the presence of air. Density fluctuations of air produce wander of
the photons beams and beam size variations that could reduce our collection
efficiency. To estimate the relevance of these effects we consider
the experimental results obtained by Resch et al. \cite{Resch_OptExpr_2005}
that performed measurements with entangled photons propagating between
two buildings on the Vienna sky at a distance of $7.8\,\mathrm{km}$.
Due to air turbulences, the authors observed large beam size variations
(displacement of the centre of the beam and changes of diameter) of
about $25\,\mathrm{cm}$ at distance $7.8\,\mathrm{km}$ from the
source. Making a linear extrapolation of the Resch et al. results
to our experimental lengths ($\approx0.8\,\mathrm{km}\ll7.8\,\mathrm{km}$),
we expect that beam size variations should be reduced to less than
$2.6\,\mathrm{cm}$ in our experiment. The diameter of lenses $L'_{A}$
and $L'_{B}$ is $18\,\mathrm{cm}$ whilst the diameter of entangled
beams impinging on lenses $L'_{A}$ and $L'_{B}$ is expected to be
$<14\,\mathrm{cm}$, then, the $2.6\,\mathrm{cm}$ displacements should
not affect appreciably the collection efficiency. 

Another effect that could reduce the collection efficiency is the
air absorption at the wavelengths of the entangled photons ($\lambda=810\,\mathrm{nm}\pm5\,\mathrm{nm}$).
However, air is known to exhibit a transmission window at these frequencies
(see Fig.8 in \cite{Gisin_RevModPhys_2002}) and, thus, losses due
to absorption are expected to be not dramatic. Precise in loco measurements
would be needed to evaluate losses due to absorption. However, one
can obtain a very rough upper estimate of the losses due to absorption
yet using the results by Resch et al. obtained with $810\,\mathrm{nm}$
entangled photons propagating on the Vienna sky. In their experiment
with the $7.8\,\mathrm{km}$ distance they observed that only a 1.4\%
fraction of the photons emitted by the source actually reached the
detectors at the $7.8\,\mathrm{km}$ distance. Note that the average
beam displacements in their experiment were very large ($25\,\mathrm{cm}$)
and greater than the diameter of their collecting telescopes. Then,
beam displacements are expected to appreciably reduce the photon counts
in their experiment. However, also assuming that losses would be entirely
due to absorption, one would obtain the absorption coefficient $\alpha=5.5\times10^{-4}\,\mathrm{m}^{-1}$
that would correspond to a 65\% collection efficiency for a beam that
propagates at a $800\,\mathrm{m}$ distance. Tacking into account
for the entangled photon intensities reported in \cite{Resch_OptExpr_2005},
we can conclude that, using a $120\,\mathrm{mW}$ laser diode and
using suitable compensation procedures and a proper optical design,
the number of measured coincidences should be $\gg10^{5}$ coincidences/s
that is more than $10^{4}$ coincidences during the acquisition time
$\delta t=0.1\,\mathrm{s}$.

\subsection{Equalization of the optical paths.}

The optical apparatus in Fig.\ref{fig:6} provides the collection
of a large number of entangled photons (99\%-fidelity) emitted within
the 0.35\textdegree{} aperture angle during the short acquisition
time $\delta t=0.1\,\mathrm{s}$. However, a high value of the lower
bound $\beta_{t,min}$ can be obtained only if also parameter $\bar{\rho}$
has a very small value, that is if the uncertainty on the equalization
of the optical paths \emph{OA }and \emph{OB} is small enough. In our
experiment, the thickness of the sensitive layer of the polarizing
filters is $220\,\mathrm{\mu m}$ and this leads to an intrinsic uncertainty
$\Delta d\lesssim220\,\mathrm{\mu m}$, that defines the minimum obtainable
value of $\bar{\rho}$ in Eq.\eqref{eq:4}. Then, it is sufficient
to require that the difference between the optical paths of the entangled
photons is much smaller than $\Delta d\lesssim220\,\mathrm{\mu m}$.
This goal will be reached using the interferometric method described
below. Initially, polarisers $P_{A}$ and $P_{B}$ will be positioned
approximatively at the same distance (within one centimeter) using
\emph{GPS} and, successively, the distances will be equalized with
the interferometric method with an estimated accuracy better that
$10\,\mathrm{\mu m}$.

\begin{figure}
\centering{}\includegraphics[scale=0.2]{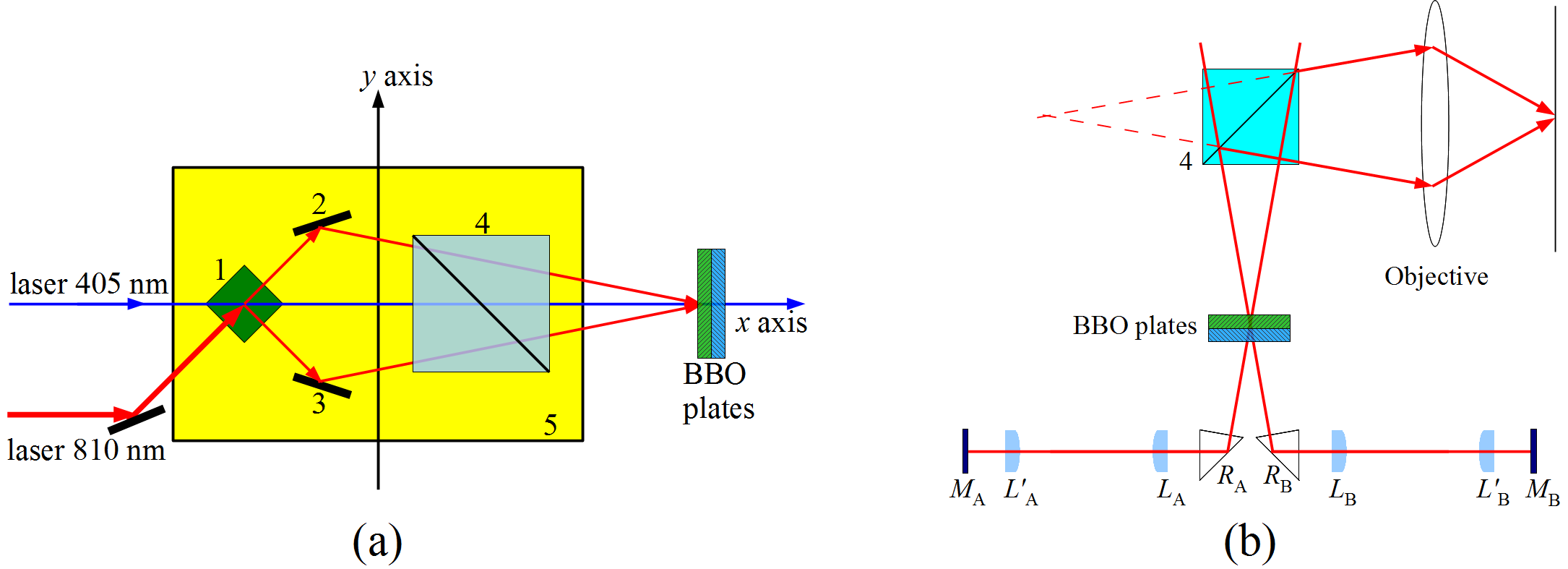}\caption{\label{fig:7} (a): system used to obtain two laser beams impinging
at $\pm2\text{\textdegree}$. 1 and 4 are beam splitters, 2,3 are
mirrors and 5 is a rigid plate that is mounted on a \emph{x}-\emph{y}
carriage that allows its insertion and removal. Beam splitter 4 needs
to collect the beams reflected by mirrors $M_{A}$ and $M_{B}$ as
shown in Fig.(b). (b): A schematic view of the interferometer. The
laser beams are reflected by mirrors $M_{A}$ and $M_{B}$, go back
and produce a interference pattern in the surface corresponding to
the central plane of the \emph{BBO} plate. Then, beams are reflected
by beam splitter 4 and a magnified image of the interference pattern
is produced by the objective. }
\end{figure}
To obtain the paths equalization, we will use a $810\,\mathrm{nm}$
laser diode with low coherence length ($0.1-0.2\,\mathrm{mm}$). Using
the simple optical system in Fig.\ref{fig:7}(a), we will obtain two
beams that impinge with equal phases on the central point of the \emph{BBO}
plates with incidence angles $\pm2\text{\textdegree}$. In such a
way the outgoing beams follow the same paths of the entangled beams.
Then, the \emph{BBO} plates will be removed using a translator and
polarisers $P_{A}$ and $P_{B}$ will be replaced by two mirrors $M_{A}$
and $M_{B}$. The laser beams reflected by mirrors $M_{A}$ and $M_{B}$
will go back and will again meet the position of the central plane
of the \emph{BBO} plates (now removed) where they form an interference
pattern with a $\approx10\,\mathrm{\mu m}$ interline. The transmitted
beams will be collected by beam splitter 4\emph{ }in Fig.\ref{fig:7}(b)
and a magnified image of fringes will be produced by the objective.
Fringes will be visible only if the difference of optical paths is
lower than the coherence length of the laser beam ($0.1-0.2\,\mathrm{mm}$)
with a maximum visibility when the optical paths are equalized. Translating
one of the mirrors with a motorized micrometer carriage and doing
real-time measurements of the fringes visibility, the optical paths
can be equalized within $\pm10\,\mathrm{\mu m}$ (for a detailed analysis
see \cite{Cocciaro_PLA_2011}). Satisfactory preliminary tests of
the proposed optical scheme have been carried out in our laboratory
over distances of about $10\,\mathrm{m}$. Once the optical paths
are equalized, mirrors $M_{A}$ and $M_{B}$ should be replaced by
polarizes $P_{A}$ and $P_{B}$, using the procedures described in
\cite{Cocciaro_PLA_2011} that ensure that the surfaces of the polarizing
layers just coincide with those of the mirrors within $\pm10\,\mathrm{\mu m}$.

\begin{SCfigure}[50]
 \centering
 \includegraphics[width=0.4\textwidth]{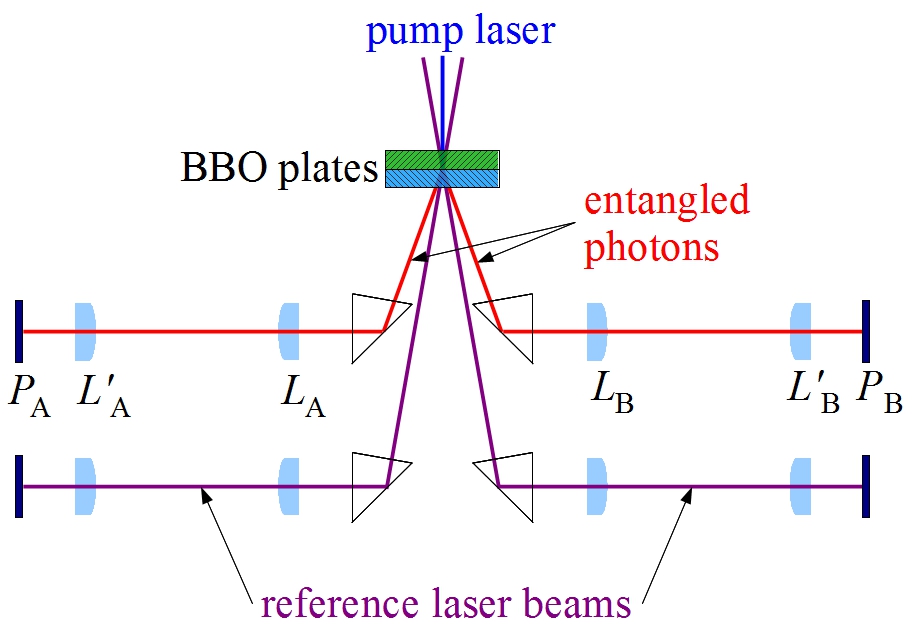}
 \hspace{0.05in}
 \caption{Schematic view of the paths of the entangled photons and of the reference laser beams.}
 \label{fig:8}
\end{SCfigure}

The above procedure to equalize the optical paths leads to a satisfactory
equalization (within $\pm10\,\mathrm{\mu m}$) at a given time. However,
due to the long optical paths characterizing our experiment, temperature
variations of the air refractive index can produce the optical paths
variation:

\begin{equation}
\Delta L_{ott}=\left(\frac{\partial n^{*}}{\partial T}\right)L_{0}\Delta T,\label{eq:6}
\end{equation}
where $L_{0}$ is the initial path length, $n^{*}$ is the group refractive
index of air at the wavelength $\lambda=810\,\mathrm{nm}$, $\nicefrac{\partial n^{*}}{\partial T}=9.47\times10^{-7}\,\mathrm{K}^{-1}$
\cite{CiddorEquation} and $\Delta T$ is the difference between the
mean temperatures along the paths of the two entangled photons. Substituting
the path length value $L_{0}=800\,\mathrm{m}$ in Eq.\eqref{eq:6},
we get $\Delta L_{ott}=0.76\,\mathrm{mm}$ for $\Delta T=1\text{\textdegree}\,\mathrm{C}$.
Other contributions are expected from changes of air pressure and
humidity, from thermal expansion and from Earth tides. The resulting
expected variations of the optical paths are much greater than the
accuracy required on the equalization of the optical paths. Then,
a suitable feedback procedure to keep constant the paths difference
is needed. To reach this goal, polariser $P_{B}$ will be placed on
a carriage driven by an electric motor and a proper feedback voltage
will be sent to the electric motor. To obtain the feedback signal
we want to build an interferometer with reference laser beams ($\lambda\approx780\,\mathrm{nm}$)
moving along the arms on paths parallel to those of the entangled
photons but spatially separated at an average distance of $20-30\,\mathrm{cm}$
(see Fig.\ref{fig:8}). Since the paths are sufficiently close, we think that the optical
paths variations due to the above effects are similar for the entangled
photons and for the reference beams. If this assumption is correct,
one can measure the paths variations of the reference beams to obtain
a feedback signal to translate the $P_{B}$ polariser. Of course,
the validity of this assumption has to be firstly checked. If the
assumption would not be verified, it would be needed to use more complex
feedback schemes as, for example, those used in \cite{Peng_PhysRevLettl_2005}
where the reference beams follow the same paths of the entangled photons.
Finally, due to the long optical paths, a special care has to be devoted
to take under control the laser beam pointing and possible drifts
of the entangled beams trajectories.

\section{\label{sec:Conclusions}Conclusions}

In this paper, we propose a long distance \emph{EPR} experiment performed
exploiting the long paths that characterize the\emph{ EGO} structures
to test the superluminal models of \emph{QM}. Such an experiment should
increase by about two orders of magnitude the actual lower bounds
for the tachyon velocities. This goal can be reached using special
compensation methods to obtain high intensity sources of entangled
photons and interferometric methods to minimize the uncertainty on
the optical paths. An important feature of this experiment is that
the measuring points \emph{A} and \emph{B} are exactly aligned along
the West-East axis. This alignment ensures that the experimental apparatus
is sensitive to any possible orientation of the velocity vector of
the preferred frame of tachyons. Therefore, if the Quantum Mechanic
correlations between entangled photons would be entirely or partially
due to superluminal communications and if the tachyon velocity in
the \emph{PF} would be lower than the $\beta_{t,min}$ value shown
in the curve \emph{III} in Fig.\ref{fig:4}, the measured correlations
should exhibit appreciable deviations from the predictions of \emph{QM}.
In such a latter case, also the actual velocity of tachyons and the
actual velocity of the \emph{PF }could be obtained from further experimental
measurements. If no deviation from the predictions of \emph{QM} would
be found, our experiment would provide a lower bound for the tachyon
velocities two orders of magnitude higher than the actual ones. In
our previous laboratory experiment \cite{Cocciaro_PLA_2011} we tested
only the simplest model of superluminal quantum communications where
all correlations between entangled photons are only due to superluminal
communications and no correlation exists if there is no communication.
In such a special case, the model can be tested orienting the polarizing
axes of polarisers $P_{A}$ and $P_{B}$ both at a $\nicefrac{\pi}{4}$
angle with respect to the horizontal direction {[}\emph{H} in Eq.\eqref{eq:1}{]}
and measuring the coincidences between photons passing through both
the polarisers. However, according to Eberhard \cite{Eberhard_1989},
this is only one of the possible superluminal models but more complex
models are also possible. In particular, some correlation between
entangled particles could be already present at the beginning (hidden
variables) and only a part of quantum correlations could be due to
superluminal communications. In such a case, also if particles have
not sufficient time to communicate, some correlations between entangled
particles remain. According to the Bell theorem, correlations due
to hidden variables alone cannot reproduce entirely \emph{QM} correlations
and, in particular, they have to satisfy the Bell inequalities. Then,
a general test of any kind of possible superluminal model can be made
only measuring a Bell-like inequality. The Bell inequalities and other
equivalent inequalities proposed in the literature, need measurements
of coincidences between entangled photons passing through polarisers
$P_{A}$ and $P_{B}$ for at least four different orientations of
their axes. In our experiment, one channel polarisers (polarizing
filters) are used and special B.C.H.S.H. inequalities must be considered
\cite{Aspect_2002}. In particular, for any hidden variables model,
it has been shown that the quantity:

\begin{equation}
M=\frac{N(a,b)-N(a,b')+N(a',b)+N(a',b')-N(a',\infty)-N(\infty,b)}{N(\infty,\infty)}\label{eq:7}
\end{equation}
must satisfy the inequality

\begin{equation}
-1\leq M\leq0.\label{eq:8}
\end{equation}
\emph{N} represents the number of measured coincidences between photons
passing through polarisers $P_{A}$ and $P_{B}$ for some different
orientations of the polarisers. \emph{a }and $a'$ are two different
orientations of polariser $P_{A}$, \emph{b }and $b'$ are two different
orientations of polariser $P_{B}$ and symbol $\infty$ corresponds
to a removed polariser. The maximum deviation of \emph{QM }correlations
from condition \eqref{eq:8} occurs if the polarization directions
\emph{a}, $a'$, \emph{b }and $b'$ are represented by angles $\theta_{a}=0\text{\textdegree},\theta_{a'}=45\text{\textdegree},\theta_{b}=22.5\text{\textdegree}$
and $\theta_{b}=67.5\text{\textdegree}$ with respect to the vertical
axis. With this choice, a positive value of \emph{M} is predicted
by \emph{QM} in disagreement with inequality \eqref{eq:8}.

From Eq.\eqref{eq:7} we see that 7 different measurements of coincidences
have to be performed to verify inequality \eqref{eq:8}. The measurements
will be performed in this way: polarisers $P_{A}$ and $P_{B}$ will
be oriented along the first combination (\emph{a},\emph{b}) occurring
in Eq.\eqref{eq:7} and coincidences will be detected during an entire
sidereal day, then the polarisers orientations will be changed to
the second combination (\emph{a},$b'$) and measurements will be repeated
until all the seven combinations have been considered. Finally, the
value of quantity \emph{M} at the different sidereal times \emph{t
}will be obtained substituting in Eq.\eqref{eq:7} the coincidences
measured at the same sidereal times of different sidereal days. In
such a way, our experiment will provide a complete test of any possible
model of \emph{QM }superluminal communications.

\appendix

\section*{Appendix}

\setcounter{section}{1}

The Alice (\emph{A}) and Bob (\emph{B}) polarization measurements
will be represented by the space-time events $\left(\vec{x}_{A},ct_{A}\right)$,
$\left(\vec{x}_{B},ct_{B}\right)$ in the laboratory frame \emph{R}
and by $\left(\vec{x}'_{A},ct'_{A}\right)$, $\left(\vec{x}'_{B},ct'_{B}\right)$
in the tachyons preferred frame \emph{$R'$}, respectively. The invariance
of the space-time intervals writes

\begin{equation}
\left(d_{AB}\right)^{2}-\left(\Delta ct\right)^{2}=\left(d'_{AB}\right)^{2}-\left(\Delta ct'\right)^{2}\label{eq:A1}
\end{equation}
where $d_{AB}=\left|\Delta\vec{x}\right|$, with $\Delta\vec{x}=\vec{x}_{B}-\vec{x}_{A}$,
$\Delta ct=ct_{B}-ct_{A}$, $d'_{AB}=\left|\vec{x}'_{B}-\vec{x}'_{A}\right|$
and $\Delta ct'=ct'_{B}-ct'_{A}$. Dividing both sides of Eq.\eqref{eq:A1}
by $\left(\Delta ct'\right)^{2}$ we get
\begin{equation}
\left(\frac{d'_{AB}}{\Delta ct'}\right)^{2}=1+\frac{d_{AB}^{2}-\left(\Delta ct\right)^{2}}{\left(\Delta ct'\right)^{2}},\label{eq:A2}
\end{equation}
where $\Delta ct'$ satisfies the Lorentz equation: 
\begin{equation}
\Delta ct'=\gamma\left(\Delta ct-\vec{\beta}\cdot\Delta\vec{x}\right),\label{eq:A3}
\end{equation}
with $\vec{\beta}=\nicefrac{\vec{V}}{c}$ ($\vec{V}$ is the velocity
of $R'$ with respect to \emph{R}) and $\gamma=\nicefrac{1}{\sqrt{1-\beta^{2}}}$.
Substituting $\Delta ct'$ of Eq.\eqref{eq:A3} in the right-hand
side of Eq.\eqref{eq:A2} we obtain 
\begin{equation}
\left(\frac{d'_{AB}}{\Delta ct'}\right)^{2}=1+\frac{\left(1-\beta^{2}\right)\left[1-\left(\frac{\Delta ct}{d_{AB}}\right)^{2}\right]}{\left(\frac{\Delta ct-\vec{\beta}\cdot\Delta\vec{x}}{d_{AB}}\right)^{2}}.\label{eq:A4}
\end{equation}
If $v_{t}=\beta_{t}c$ is the tachyons velocity in $R'$, the quantum
communication cannot occur (the Bell inequalities are verified) only
if
\begin{equation}
\beta_{t}<\frac{d'_{AB}}{\left|\Delta ct'\right|}.\label{eq:A5}
\end{equation}
Inequality \eqref{eq:A5} would be always verified for any finite
$\beta_{t}$ provided the two conditions 
\begin{equation}
\Delta ct=0\:\:\:\:\:\wedge\:\:\:\:\:\vec{\beta}\cdot\Delta\vec{x}=0\label{eq:A6}
\end{equation}
would be satisfied that lead to $\frac{d'_{AB}}{\left|\Delta ct'\right|}\rightarrow\infty$
in Eq.\eqref{eq:A4}. However, due to the experimental features, conditions
\eqref{eq:A6} can be only approximatively satisfied and, thus, discrepancies
with the \emph{QM} will occur only if 
\begin{equation}
\beta_{t}<\beta_{t,min},\label{eq:A7}
\end{equation}
where $\beta_{t,min}$ is the minimum value of $\frac{d'_{AB}}{\left|\Delta ct'\right|}$
in Eq.\eqref{eq:A4} that can be effectively reached in the experiment.
The basic experimental features are the uncertainty $\Delta d$ on
the equality of the optical paths, that leads to the condition 
\begin{equation}
-\Delta d\leq\Delta ct\leq\Delta d,\label{eq:A8}
\end{equation}
and the acquisition time $\delta t$ that leads to the condition
\begin{equation}
-\beta\left|\Delta\vec{x}\right|\sin\chi\sin\frac{\pi\delta t}{T}\leq\vec{\beta}\cdot\Delta\vec{x}\leq\beta\left|\Delta\vec{x}\right|\sin\chi\sin\frac{\pi\delta t}{T},\label{eq:A9}
\end{equation}
where $\chi$ is the angle between vector $\vec{V}$ and the polar
axis while \emph{T} is the sidereal day. Eq.\eqref{eq:A9} is obtained
(without lack of generality) using the expressions $\Delta\vec{x}=\left|\Delta\vec{x}\right|\left(1,0,0\right)$
and $\vec{\beta}=\beta\left(\sin\chi\cos\frac{2\pi t}{T},\sin\chi\sin\frac{2\pi t}{T},\cos\chi\right)$
that lead to 
\begin{equation}
\vec{\beta}\cdot\Delta\vec{x}=\beta\left|\Delta\vec{x}\right|\sin\chi\cos\frac{2\pi t}{T}.\label{eq:A10}
\end{equation}
From Eq.\eqref{eq:A10} we infer that $\vec{\beta}\cdot\Delta\vec{x}$
is minimized if the acquisition time interval $\delta t$ is centred
on times $t_{1}=\frac{T}{4}$ or $t_{2}=\frac{3}{4}T$ that is if
$t_{1,2}-\frac{\delta t}{2}\leq t\leq t_{1,2}+\frac{\delta t}{2}$
. Then, in both cases $\vec{\beta}\cdot\Delta\vec{x}$ in Eq.\eqref{eq:A10}
will satisfy conditions \eqref{eq:A9}. The minimum value of $\left|\frac{d'_{AB}}{\Delta ct'}\right|$
in Eq.\eqref{eq:A4} ($\beta_{t,min}$ in Eq.\eqref{eq:A7}) occurs
if ratio $\nicefrac{1-\left(\frac{\Delta ct}{d_{AB}}\right)^{2}}{\left(\frac{c\Delta t-\vec{\beta}\cdot\Delta\vec{x}}{d_{AB}}\right)}{}^{2}$
is minimized with $\Delta ct$ and $\vec{\beta}\cdot\Delta\vec{x}$
satisfying conditions \eqref{eq:A8} and \eqref{eq:A9}. It can be
easily seen that this occurs if $\left(\Delta ct\right)^{2}=\left(\Delta d\right)^{2}$
and $\left(c\Delta t-\vec{\beta}\cdot\Delta\vec{x}\right)^{2}=\left(\Delta d+\beta\left|\Delta\vec{x}\right|\sin\chi\sin\frac{\pi\delta t}{T}\right)^{2}$
. Then, we find 
\begin{equation}
\beta_{t,min}=\sqrt{1+\frac{\left(1-\beta^{2}\right)\left(1-\bar{\rho}^{2}\right)}{\left(\bar{\rho}+\beta\sin\chi\sin\frac{\pi\delta t}{T}\right)^{2}}},\label{eq:A11}
\end{equation}
 where $\bar{\rho}\equiv\frac{\Delta d}{d_{AB}}$. Eq.\eqref{eq:A11}
just corresponds to Eq.\eqref{eq:4}.

\bibliographystyle{iopart-num}
\bibliography{ArticoloFinaleCongressoCastiglioncello2012}

\end{document}